\documentclass[showpacs,twocolumn,prl]{revtex4}
\usepackage{amsmath,graphicx}

\begin{document}
\title{ Magnetic field resistant quantum interferences in bismuth nanowires based Josephson junctions}
\author{Chuan Li$^{1}$, A. Kasumov$^{1,5}$, A. Murani$^{1}$, Shamashis Sengupta$^{2}$, F. Fortuna$^{2}$, K. Napolskii$^{3,4}$, D. Koshkodaev $^{4}$, G. Tsirlina$^{3}$, Y. Kasumov$^{5}$, I. Khodos$^{5}$, R. Deblock$^{1}$, M. Ferrier$^{1}$, S. Gu\'eron$^{1}$ and H. Bouchiat$^{1}$}%, K. Napolskii $^{5}$, P. Ioselevich$^{5}$ and  M. Feigelman$^{5}$ }

\affiliation{$^{1}$ LPS, Univ. Paris-Sud, CNRS, UMR 8502, F-91405 Orsay Cedex, France}
\affiliation{$^{2}$ CSNSM, Univ. Paris-Sud, IN2P3, UMR 8609, F-91405 Orsay Cedex, France}
\affiliation{$^3$ Faculty of Chemistry, Moscow State University, Leninskie Gory, 1-blg.3, Moscow, 119991, Russia}
\affiliation{$^4$ Department of Material Sciences, Moscow State University, Leninskie Gory, Moscow, 119991, Russia }
\affiliation{$^5$ Institute of Microelectronics Technology and High Purity Materials, RAS, ac. Ossipyan, 6,  Chernogolovka, Moscow Region, 142432, Russia}

\begin{abstract}
We investigate proximity induced superconductivity in micrometer-long bismuth nanowires connected to superconducting electrodes with a high critical field.  At low temperature we measure a supercurrent that persists in magnetic fields as high as the critical field of the electrodes (above 11 T). The critical current  is also strongly  modulated by the magnetic field.  In certain samples we find regular, rapid SQUID-like periodic oscillations  occurring up to high fields. Other samples exhibit less periodic but full modulations of the critical current on Tesla field scales, with field-caused extinctions of the supercurrent. 
 These findings indicate the existence of  low dimensionally, phase coherent, interfering conducting regions through the samples, with a subtle interplay between orbital and spin contributions.
We relate these surprising results to the electronic properties of the surface states of bismuth, strong Rashba spin-orbit coupling,  large effective g factors, and their effect on the induced superconducting correlations.

73.63.-b, 74.45.+c, 74.78.-w, 74.78.Na
  
\end{abstract}

\maketitle

In the superconducting proximity effect, singlet pair correlations can penetrate quite far  (on the micron scale) into a non superconducting (normal) conductor. This penetration, that can lead to  supercurrents through normal conductors several micrometers long connected to two superconductors,  results from quantum interference between all conduction channels in the sample. In a microscopic picture, the supercurrent is carried by Andreev states,  combinations of time reversed electron and hole  wavefunctions confined to the normal conductor. It is thus natural to consider that this interference is destroyed not only by inelastic scattering,  but also by time reversal symmetry breaking. Indeed, a magnetic field is known to suppress the supercurrent via both orbital (Aharonov Bohm phase accumulation) and spin (Zeeman dephasing) effects.  Nevertheless, supercurrents have been induced through ferromagnets. The oscillatory sign and decaying intensity of the supercurrent with increasing ferromagnet thickness is an illustration of the dephasing role played by the exchange field. On the other hand,  the time reversal invariant   spin orbit interactions,  by  imposing strong correlations between spatial and spin components of the induced  Andreev pairs, offer new possibilities such as  coupling between singlet and triplet pairing \cite{gorkov, bergeret}, arbitrary Josephson phase shifts in  an exchange or a Zeeman field ($\phi$ junction behavior) \cite{buzdin,feinberg,nazarov} and the possible formation of Majorana fermions at the interface between semiconducting nanowires and  superconducting electrodes \cite{frolov}.

In this Letter, we probe the superconducting proximity effect in crystalline bismuth nanowires, a system with extremely high  Rashba spin orbit coupling (SOC), connected to superconducting electrodes with standard s-wave pairing and a very  high critical field $H_c$. 
The complex interference pattern we measure, up to magnetic fields such that the Zeeman energy, $E_Z$, becomes of the order of the spin-orbit and Fermi energies ($E_F$), uniquely reveals the role played by both spin and orbital degrees of freedom.

Bismuth is a semi-metal with rhombohedral structure whose bulk electronic properties have been extensively studied: three barely filled  electron bands coexist with a single, nearly filled,  heavy hole band. Bi\rq{}s strong atomic spin orbit energy leads to extremely high effective g factors ($g_{eff}\sim100$), that depend on the applied magnetic field direction. Moreover the semi-metallic character leads to unusually large Fermi wavelengths, $\lambda_F\sim50~nm$ \cite{hofman}.  Therefore in nanostructures only a few $\lambda_F$  thick or wide, because of quantum confinement, the surface states rather than the bulk states should play a major role, in particular in the transport properties \cite{nikolaeva,ning}. Angle-resolved photoemission (ARPES) revealed electronic surface states  in Bi with almost free electronic mass and nanometer-size $\lambda_F$. These states are remarkable in that the energy bands display a huge Rashba spin splitting, because of the loss of inversion symmetry at the surface combined  with Bi\rq{}s high atomic SOC.  ARPES  of differently oriented Bi surfaces \cite{hofmanarpes} and spin resolved ARPES \cite{Hirahara} have found  spin  splitting energies of about 0.1 eV, as high as $E_F$. The (111) surface, perpendicular to the rhombohedral axis, is particular because it possesses states on the top bilayer, that are decoupled from the bulk states. %(the Bi atoms in the bilayer are closer to each other than to the next layer).
 One dimensional quantum spin Hall states have even been predicted at the edge of these (111) surfaces \cite{murakami}. Quite recently, scanning tunneling microscopy have indeed found 1D edge states around single crystalline bilayer  islands on the top of  BiSe or bulk Bi(111) crystals \cite{STM,Yazdani}.  1D topological states of (114) surfaces  have also been seen by ARPES \cite{Bi114}.
Thus 100 nm-wide Bi nanowires seem ideal to investigate the effect of SOC on the superconducting proximity effect, in a regime barely explored up to now, in which the spin-splitting energy of carriers is comparable to $E_F$. 
In addition, the surface states \rq{} relatively high  $g_{eff}$ (between 10 and 100, depending on the surface orientation with respect to the magnetic field) \cite{Seradjeh}, imply that $E_Z$ can also reach $E_F$ and spin splitting energy at fields on the 10 T scale. 
Finally, the relative directions between the Zeeman and orbital fields can be varied, leading to even richer physics.
 %depending also on the direction of the magnetic with respect to the considered surface.

In the following, we present experiments on Bi nanowires connected to high $H_c$ superconducting electrodes  that show striking differences with ordinary Josephson SNS junctions: the supercurrent  persists up to magnetic fields as high as the $H_c$ of the electrodes (11 T). In some cases the field even enhances the  critical current $I_c$. In addition, we find oscillations of $I_c$:  some  samples display both a periodic squid-like oscillation with a period in the hundred Gauss range and a higher, Tesla range,  modulation,  while in other samples only a strong high field modulation is found, with the  complete extinction of  $I_c$ at specific fields.  These findings point to the existence   of interfering Andreev pairs  confined to a small number of conducting regions of low dimensionality. A subtle interplay between orbital and spin contributions  is also required to explain the extent and period of interference. We discuss these unusual results in view of the properties of Bi\rq{}s surface states, the very strong Rashba SOC, and  high anisotropic  $g_{eff}$.
%\begin{figure}
  %   \centering
   %\includegraphics[width=\linewidth]{Fig1.pdf}
     %\caption{ 
     %(Top) Transmission electron microscopy   pictures of the Bi nanowires  fabricated in the same conditions than the one investigated, revealing  a small number of cavity like defects.  High resolution investigations indicate   that the monocrystalline structure  is not affected by these defects and  the absence of grain boundaries  as confirmed by electron diffraction measurements (Inset). (Bottom) Scanning electron microscopy  pictures of Bi nanowires $Bi_1$ and $Bi_3$ connected by W wires under the FIB for transport measurements, the insulating polycarbonate layer surrounding the layer is clearly visible.} 
      %\label{fig1} 
     %\end{figure}
		
The Bi nanowires are electrochemically grown in the $90\pm 10~nm$-wide pores of a polycarbonate track-etched membrane, and released by dissolution of the membrane (supplementary materials).   X-ray diffraction and  transmission electron microscopy demonstrate the high crystallinity of the few-$\mu m$-long nanowires, with no high angle grain boundaries. An approximately 10 to 20 nm-thick external amorphous layer  is also found, probably a protective residual polycarbonate coating. %are nearly mono-crystalline   %High resolution  transmission  electron microscopy and electron diffraction of such crystallites are shown in Fig. 1.  Hollow  cavity defects are also visible at the wire edges, leading to local constrictions.  These defects are not positioned at the (rare) grain boundaries. Lower resolution scanning electron micrographs of the connected nanowires we measured also reveal such cavities and constrictions the samples (see fig. 1).
Given the nm-size $\lambda_F$ of Bi\rq{}s surface states \cite{hofman}, more than 100 conduction channels are expected at the surface of the wires.  The nanowires  are most likely 
faceted polyhedra, with each facet having a potentially different crystalline orientation. Thus some facets can be insulating \cite{murakamiPRB} while others,  such as (111)  or (114) facets, may have very specific conduction properties.
The nanowires are deposited onto an oxidized Si substrate with prepatterned electrodes. The superconducting contacts to the Bi nanowires, and connection to the electrodes, are realized in a dual electron and ion beam microscope equipped with a gas injection system: the focused Ga ion beam (FIB)  decomposes a tungsten carbonyl vapor, producing a carbon and gallium-doped amorphous tungsten wire roughly 100 nm thick and wide.  The superconductive properties of these wires are  impressive, with a transition temperature $T_c \sim 4~K$,  $I_c\sim100~\mu A$, and $H_c$ above 11 T at low temperature (LT) \cite{kasumov05}.  The superconducting gap measured by scanning tunneling spectroscopy \cite{viera} is $\Delta = 0.8$ meV.  We have checked that SNS junctions with $\mu m$-long Au wires contacted in this way behave similarly to more conventionally fabricated SNS junctions \cite{chiodi}.  Because the FIB can be used to etch the Bi wire and  coating at the contact before  W deposition, this  technique ensures a   good, albeit not perfect, transparency. %In contrast,  no good contact to these Bi nanowires could be achieved using standard e-beam lithography techniques. % Scanning Electron Microscopy (SEM)  of the measured samples reveal an irregular polymer- like coating   around the nanowires,  which we attribute to incompletely dissolved carbonate membrane, that protects the bismuth   from oxidation.  % We cannot however completely exclude the formation of an amorphous oxide layer as already reported on Bi whiskers \cite{chan}.  
%The distance between the electrodes is kept above one micron to prevent shortings due to %the formation of  parallel tungsten conduction paths on the substrate. SEM  also  confirmed the absence of W contamination of the substrate beyond 300 nm away from the electrodes. 
The contacts degrade with time, so the samples were cooled within hours of their connection, except for one sample (Bi3, see below) that was kept several weeks in vacuum at 300K after the first set of measurements, and whose resistance doubled.

%The contact quality  degrades with time, as demonstrated by the steady resistance increase of the samples kept at room temperature. We attribute this to the oxidation of Bi in the contact region. We thus cooled the samples to liquid helium temperature within a few hours after their connection, except for one sample (see below) which was kept several weeks at room temperature (under vacuum) after a first series of low temperature measurements. The resistance then increased twofold.

\begin{figure}
     \centering
   \includegraphics[width=\linewidth]{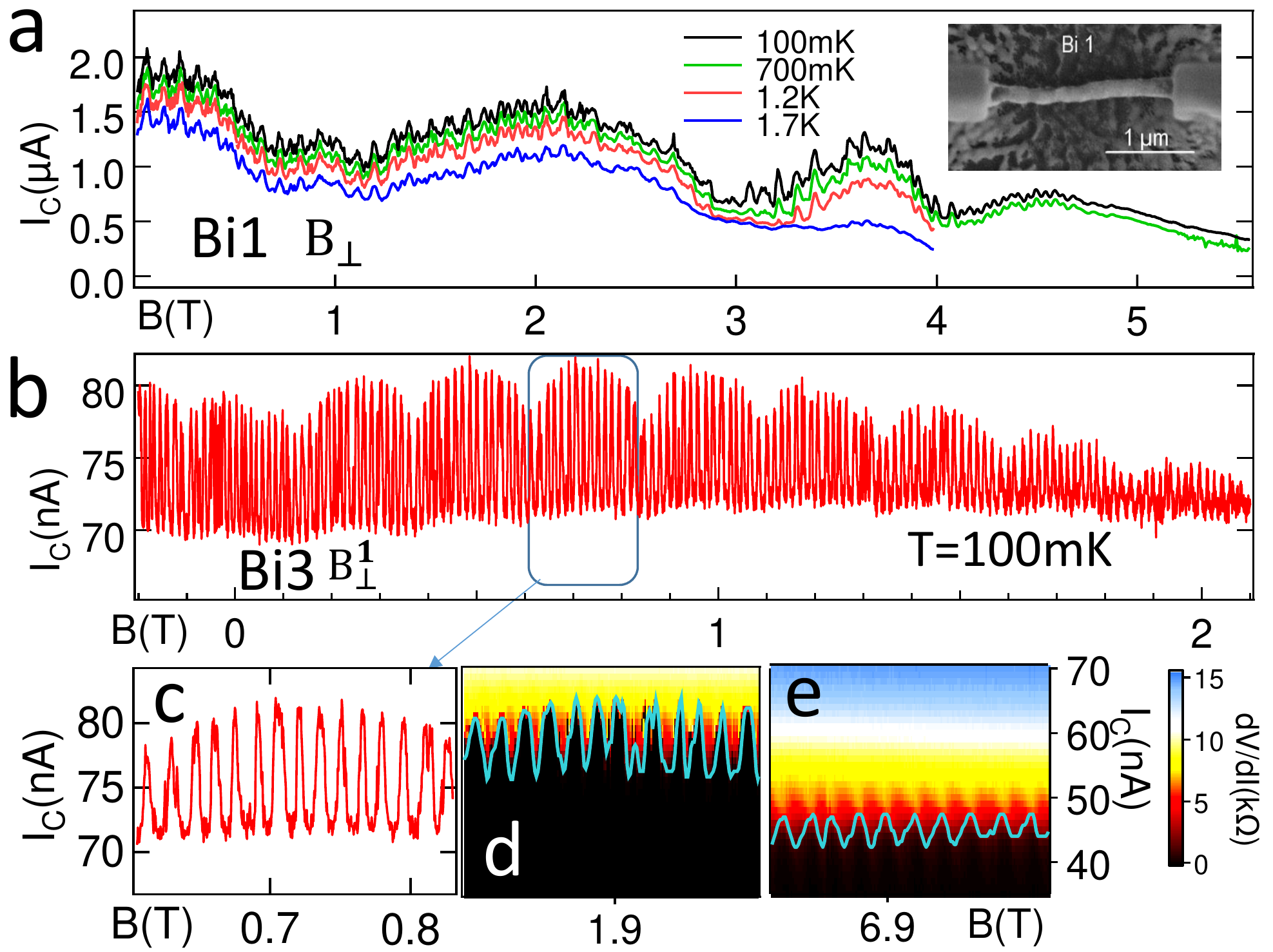}
     \caption{Magnetic field dependence of the supercurrent of $Bi_1$ (a) and $Bi_3$ (b-e), in a perpendicular field. Fast, squid-like oscillations with periods of 800 and 150 G for $Bi_1$ and $Bi_3$ respectively, are noticeable, up to unusually high fields (at least 6 T for $Bi_1$,  10 T for $Bi_3$). $Bi_3$ displays an additional periodic modulation with a 2300 G period, and an irregular modulation of $Bi_1$\rq{}s critical current occurs on the Tesla scale.  Two kinds of  switching measurements were performed on $Bi_3$, with different time scales. As expected, the faster measurements (b) and (c) yield somewhat higher switching currents  than the slow measurements (d) and (e). Inset: Scanning electron micrograph of $Bi_1$ and its superconducting contacts.}
      \label{fig2}
     \end{figure}
		
\begin{figure}
     \centering
   \includegraphics[width=\linewidth]{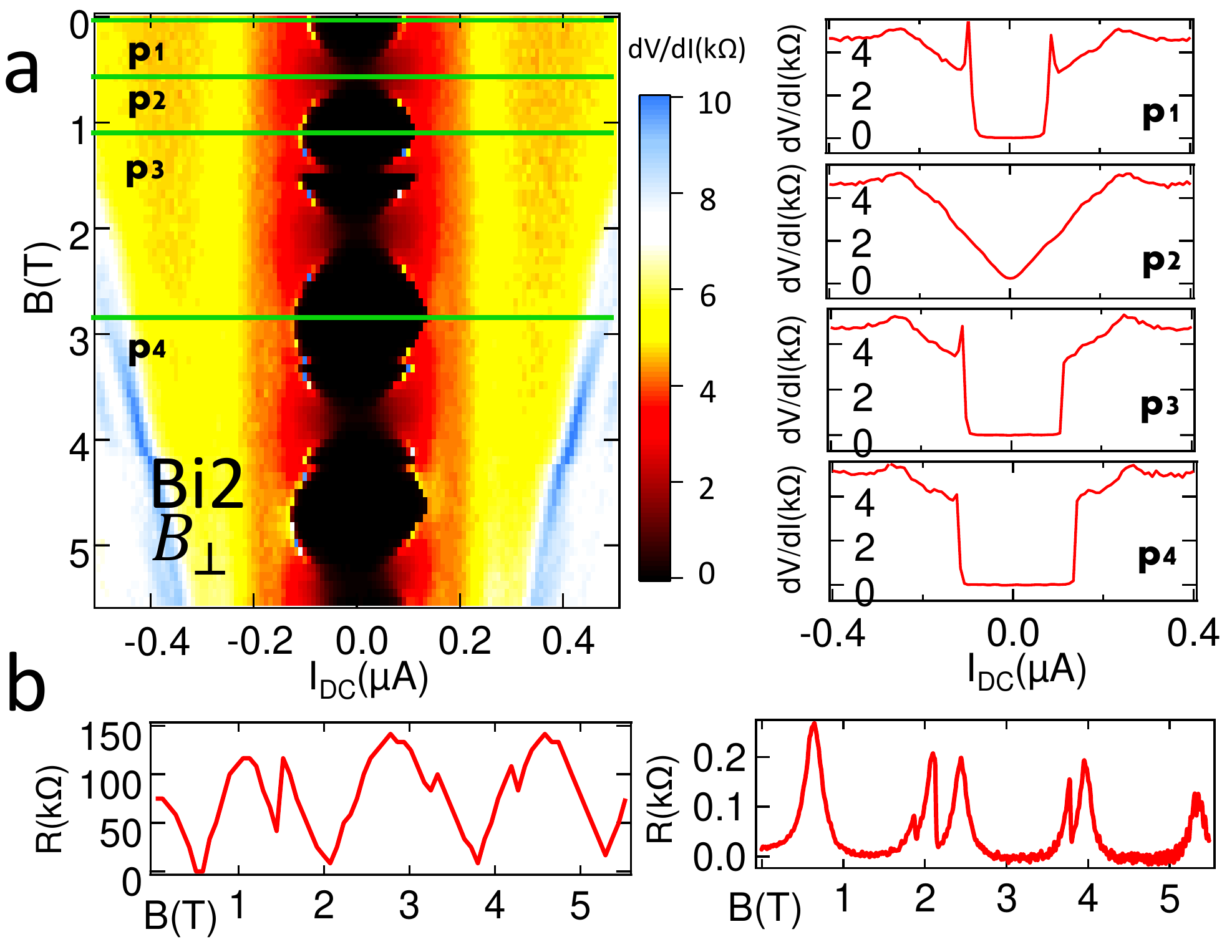}
     \caption{ (a) Color-coded differential resistance of  $Bi_2$, as a function of dc current and magnetic field, with some characteristic differential resistance curves on the right. (b) and (c) Field dependence of $I_c$ and zero bias differential resistance extracted from (a). Note the oscillatory behavior on the Tesla scale, and also how the maximal critical current increases with field.}
      \label{fig3}
     \end{figure}
		
We have investigated the LT resistance of ten such samples. Below the $T_c$ of the W electrodes, the resistance is mostly due to the  two probe resistance of the Bi nanowires. Although the wires have similar dimensions, this resistance varies widely, between 1 and 30 k$\Omega$. Since the intrinsic resistance of the Bi wires is only expected to be few hundred $\Omega$ (if one extrapolates reports on much longer wires of similar diameters \cite{nikolaeva}), this indicates that the wire/contact interface resistance dominates. 

Proximity induced superconductivity gives rise to a resistance decrease below the $T_c$ of the  W electrodes in five samples out of ten. A supercurrent,  corresponding to a zero resistance state, is detectable in three samples. Two other samples display an incomplete proximity effect: the resistance drop is small (3 to 10 percent), and turns into a resistance increase (of about 10 percent) as the  temperature is lowered further. The LT differential resistance of those two samples is  peaked at low current, due to the interplay of interactions and a  low transparency of the Bi/W interface \cite{morpurgo}. These results are an indication that our Bi nanowires are not {\it intrinsically} superconducting, in contrast to the superconductivity below 1 K found in prior work on Bi nanowires \cite{chan}. Those nanowires were  unprotected against oxidation,  resulting in more pronounced  surface disorder than ours. Kobayashi et al. also found intrinsic superconductivity  with $T_c= 8~K$ and $H_c=4~T$  in highly  disordered nanowires with nm-size grains, but no intrinsic superconductivity down to 0.5 K in oxide-free cristalline Bi nanowires \cite{kobayashi}. 
There is however one report of intrinsic superconductivity in arrays of single crystal Bi nanowires grown, as ours, in polycarbonate membranes (but with a different electrolyte) \cite{ye08}, with $T_c\sim 0.64~K$ and $H_c$ smaller than 0.5 T. 

\begin{figure}
     \centering
   \includegraphics[width=\linewidth]{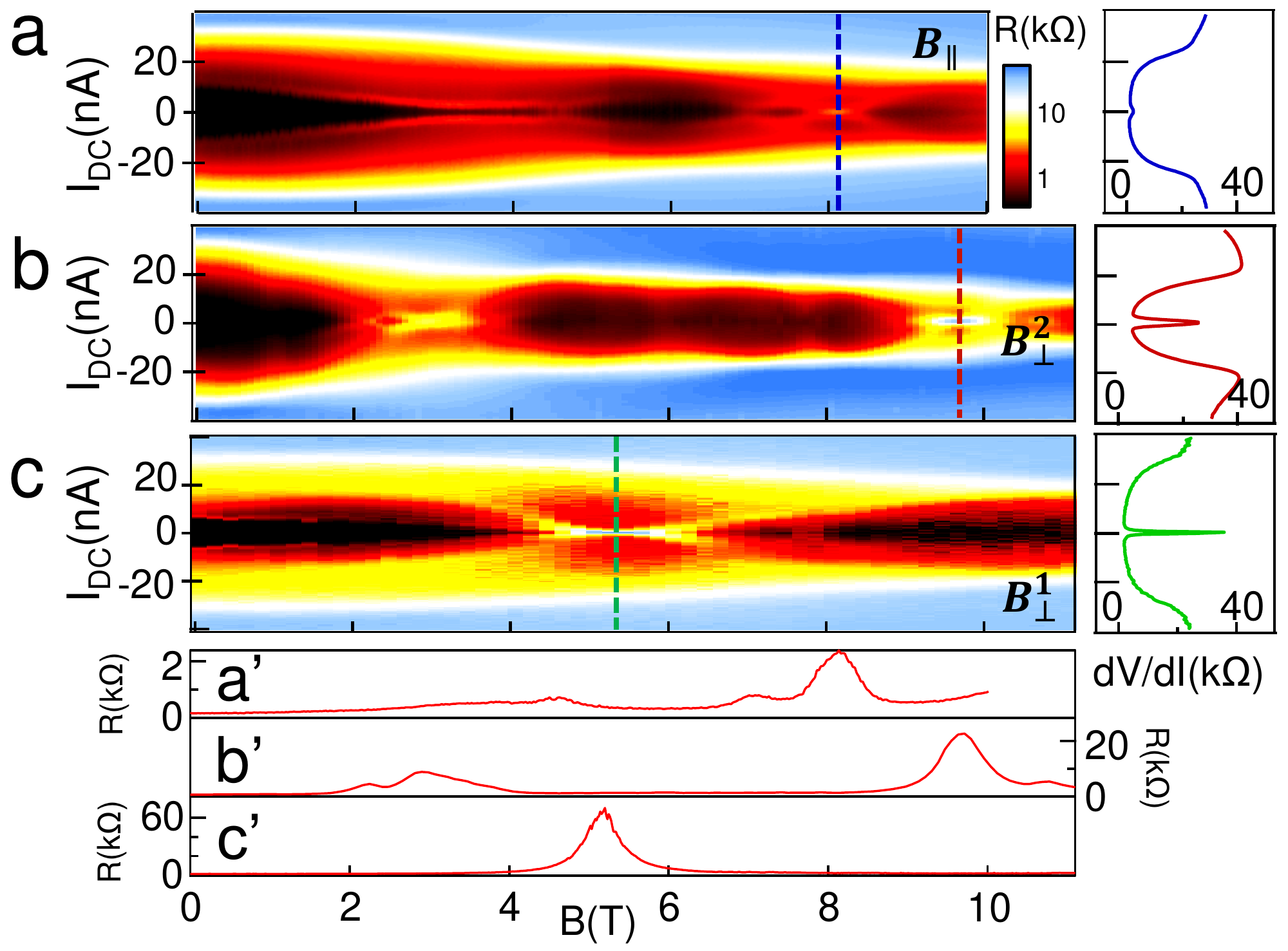}
     \caption{ Color-coded differential resistance as a function of current and field, and extracted curves, for  3 field orientations, on sample $Bi_3^*$ (that corresponds to sample $Bi_3$ after thermal cycling and aging).  (a) field along the nanowire axis; (b), (c) two field orientations perpendicular to the wire axis. 
      }
      \label{fig1}
     \end{figure}
		
In the following we focus on the 3 samples with a detectable supercurrent,  $Bi_1$, $Bi_2$ and $Bi_3$, all of which are 2$\mu m$ long. Their normal state resistances $R_N$, measured below the W wires\rq{} $T_c$,  are respectively  1, 13 and $16~k\Omega$. We also present successive cool downs of $Bi_3$, with changed orientations between wire and field. The sample is called $Bi_3^*$ since its room temperature resistance increased to $27~k\Omega$, implying a worsening of the contact to the W electrodes.
The zero bias resistance drops to zero below 0.8 K, and the differential resistance curves display a zero resistance state for currents below a switching current of  $1.5, 0.1$, and $0.075~\mu A$  for  $Bi_1$,  $Bi_2$ and  $Bi_3$ respectively,  in zero magnetic field and  at 100 mK. In the following, we equate this switching current and $I_c$. We also extrapolate $I_c=30 nA $ for   $Bi_3^*$  even though it does not display a fully zero  resistance state. 
The $R_N I_c$ product ranges between  0.75 and 1.5  meV, the same order of magnitude as the superconducting gap of the W wires. This is consistent with what is expected of short  Josephson SNS junctions with an induced gap of the order of the gap of the electrodes.  The short junction behavior implies that the superconducting gap is less than 10 times the Thouless energy $ \hbar D/L^2~ \sim \hbar v_Fl_e/L^2 $ \cite{dubos}. A Fermi velocity $v_F \leq  3~10^5$ m/s yields a mean free path $l_e\simeq 2\mu m \simeq L$, which confirms  the ballistic nature of  transport through the wires.  The temperature dependence of $Bi_3$\rq{}s differential resistance, and the Shapiro steps under irradiation at frequency f, that appear at the expected dc voltages $2eV_n= nhf$, are shown in supplementary materials. %as well as for a plain W wire. 

Our most striking result concerns the  magnetic field dependence of $I_c$ (see Figs. 1 and 2, with the field perpendicular to the wires, and Fig. 3 with  3 different orientations). %For all samples %except $Bi_3^*$
% the magnetic field was perpendicular to the nanowire axis. 
First, we find that the  supercurrent persists up to very high fields:  higher than 6 T for $Bi_1$ and $Bi_2$, and 11  T for $Bi_3$: in all cases those values are merely limited by the highest field achievable with the superconducting magnets used in the experiments. 
Second, $I_c$ of all three samples is strongly modulated by the magnetic field: two samples,  $Bi_1$ and  $Bi_3$, display SQUID-like oscillations of $I_c$, with a period of 800 G for  $Bi_1$ and  140 G for  $Bi_3$.
%, compatible with a flux quantum through the area perpendicular to the magnetic field. 
These rapid oscillations persist to high fields, up to 10 T for  $Bi_3$ (not shown).

The critical current of $Bi_3$ is also modulated with a second period of about 0.3 T.
Finally, $I_c$ is also modulated   aperiodically  on the Tesla scale for $Bi_1$, $Bi_2$ and $Bi_3^*$ (see Figs. 1, 2 and 3).  In samples without rapid SQUID-like oscillations, the high field modulation  causes a full extinction of the supercurrent in $Bi_2$ and $Bi_3^*$,  with entire magnetic field intervals having zero supercurrent and finite resistance. We also explored 3 perpendicular field  orientations for $Bi_3^*$, including one along the wire axis (Fig. 3). The field modulation patterns of $I_c$ differ. High resistance peaks occur at different fields (8, 9 and 5 T, see Fig. 3).  The small-period, squid-like oscillations of the first cool-down are not detectable in these subsequent cool downs. 
%, probably because of worsening of the contact upon thermal cycling: it is possible that one of the two conduction channels whose interference causes the supercurrent oscillation is no longer connected to the contact.

We now discuss these complex interference patterns by considering field-induced phase shifts of the Andreev pairs wave functions,  whose  origin,   involve either orbital or spin degrees of freedom. We first recall the generic field dependence of $I_c$ of ordinary  SNS junctions, with a  very high number of conduction channels. $I_c$ is strongly suppressed by magnetic field  due to two different pair breaking mechanisms. In the semi-classical limit ($\lambda_F$ much smaller than all sample dimensions), orbital phase breaking is due to the  Aharonov Bohm phase difference between  different Andreev pairs that follow different trajectories through the N. This orbital dephasing suppresses the supercurrent at fields corresponding to a flux quantum through the sample,  as observed experimentally e.g. in Au wires \cite{chiodi}.   In samples with a very small area perpendicular to the magnetic field, this orbital dephasing is weak, and the spin phase breaking caused by the Zeeman effect becomes noticeable. The Zeeman effect causes  a  phase difference  between the electron and hole components of a given Andreev pair, given by $E_ Z\tau/\hbar$  on a   trajectory  of length $L_t$ ( $\tau =L_t/v_F $ is the  time to cross the sample). 
Summing the contributions  of all Andreev pairs  trajectories, when the number of channels is large, yields an exponential  supercurrent suppression  with magnetic field. %on the  length scale  of $(\hbar D / E_Z)^{1/2}$%. 
%A similar suppression occurs when N is a ferromagnet, with $E_Z$ replaced by  the exchange energy. Note that this spin pair-breaking mechanism does not exist for triplet pairing.
 
All three Bi wires have an area perpendicular to the magnetic field of $2~\mu m$ by $100~ nm$, so that one flux quantum corresponds to a magnetic field of 50 G, three orders of magnitude smaller than the supercurrent extinction field found in the experiment. The persistence of supercurrents to  fields as  high as 10 T can only be understood if transport is confined to very few, quasi ballistic,  1D channels whose width should not exceed $\lambda_F$.  These channels could be located along the edges of particular facets parallel to the nanowire axis \cite{cylinder}. 
The small period (few 100 G), SQUID-like oscillations of $I_c$ in $Bi_1$ and $Bi_3$, are then the manifestation of quantum  interference between Andreev pairs belonging to two such 1D edge channels.  Such topological edge states could be those of the (111) or (114) surfaces, or of others possessing similar topological properties.  Since the period of the oscillation corresponds to the flux enclosed between  the interfering channels, the measured periods of 140 G for $Bi_3$ and 800 G for $Bi_1$ would correspond to  1D channels  along the samples axis, distant by 70 and 12 nm respectively. 
This interference pattern, that we interpret as due to the concentration of current along certain edges, recalls the recent observation of periodic oscillations of the Josephson current carried by spin Hall edge states in a 2D topological insulator connected to superconducting electrodes  \cite{Yacoby}. 
  
We now discuss the supercurrent modulations at higher  field, and argue that they are due to the difference between the electron and hole  wavevectors of the Andreev pairs. The magnetic field, via $E_Z$, shifts the wavevectors of carriers of opposite spin at the Fermi level \cite{feinberg, nazarov}.  Within linear approximation, the  phase  difference accumulated  between the electron and hole components of opposite spin along a 1D ballistic trajectory of length $L$ is \cite{ioselevitch, buzdin}   $ \delta \phi(B)= E_Z.L/(\hbar.v_F)=g_{eff}.\mu_B. B_{//} (\hbar . v_F/L)$, % where $L/v_F$ is the time of flight across the wire and 
 with $B_{//}$ the field component along the spin orbit field. (Note the exact similarity to 1D ballistic  superconducting/ferromagnetic/superconducting  (SFS) junctions, \cite{ buzdinRMP}).
 Typical Bi  surface states parameters ($v_F \simeq$3 $10^5 m/s$ and $g_{eff} =30 $) yield a characteristic modulation period in the Tesla range, so that we believe that the large field modulations of $I_c$ seen on all the samples are due to  this spin dephasing effect. The difference in actual interference pattern of the various samples, as well as for the three field orientations of $Bi_3^*$, is  expected, given the anisotropy of the Bi facets and the corresponding different $g_{eff}$, that can vary by more than an order of magnitude. It is easy to reproduce the experimental data on $Bi_3$ by considering the interference  between two channels of  transmissions differing by a factor 8, enclosing a surface of the order of the sample area. One has to take $g_{eff}\sim 100$ for the weakly transmitting channel, yielding an amplitude modulation of the small-period orbital SQUID-like oscillations with a 0.3 T period, and a much smaller $g_{eff}$ for the strongly transmitting channel \cite{mironov}.

In this picture, the full extinction of the supercurrent at nearly periodic field values is attributed to the Zeeman-induced $(2n+1)\pi$ phase differences (with $n$ integer) between the electron and hole components of the supercurrent-carrying Andreev pairs. Such full extinction (complete destructive interference) is thus restricted to a single current-carrying channel. This seems to be the case in $Bi_2$ and $Bi_3^*$, since they do not display SQUID-like oscillations (that require two channels). 
The $Bi_3^*$ behavior is especially dramatic around 5 T (Fig. 3 c and c'), with a zero bias resistance that peaks at a value even higher than the normal state resistance. A final important finding is the enhancement of $I_c$ by the magnetic-field. $Bi_2$\rq{}s critical current at 5 T is twice that of zero field (Fig. 2a and b).  A similar but smaller increase  between 0 and 0.75 T is also seen in $Bi_3$ (Fig. 1b). This increase of $I_c$ with field may be attributable to the strong SOC,  as predicted in $\phi$ junctions % spin orbit coupling-induced superposition of singlet and triplet pairing \cite{gorkov}, since it has been shown that high magnetic field enhances inhomogeneous superconductivity 
 \cite{buzdin, mironov}. 

We have shown evidence of quantum interference  in Bi  nanowires-based Josephson junctions, that persist up  to very high magnetic fields.  Sample dependent  periodic oscillations of the critical current reveal complex interference patterns involving both orbital and spin degrees of freedom between a small number of strongly confined 1D channels, possibly located at the edges between facets of different crystalline orientations along the wires. 
 The physical origin of  this confinement of induced superconductivity in such quasi one-dimensional channels is not yet well understood. % In particular, we hope to be able  in future studies on single crystals of known orientation, to disentangle the role played by the crystalline structure of the wires from the  existence of a high spin orbit coupling. 
One possibility is that the confinement is favored in the superconducting state by the high magnetic field, that is known to induce inhomogeneous superconductivity in 2D superconductors with large  Rashba SOC \cite{Barzykin}.

 We acknowledge  fruitful discussions with  Alexei Chepelianskii,  Michael Feigelman, Pavel Ioselevich, Pascal Simon, Sacha Buzdin  and Sergey Mironov.  This work is supported by a Franco-Russian  RFBR grant  13-02-91058-CNRS. The LPS group benefits from financial support from CNRS and  of the French national research agency (ANR SUPERGRAPH and MASH).


\begin{thebibliography}{99}
\bibitem{cylinder}An alternative hypothesis could be that the nanowires behave as nearly perfect  ballistic cylinders, with a quasi-full cancellation of the flux of a perpendicular magnetic field. But we doubt this ideal picture can realistically describe our samples, see \cite{ioselevitch}.
\end{thebibliography}
\end{document}